\documentclass[journal, 10pt]{IEEEtran}

\usepackage{blindtext}
\usepackage{relsize}
\usepackage{amsmath}
\usepackage{amsmath}

\usepackage{amssymb}
\usepackage{amsthm}
\usepackage{amsfonts}
\usepackage[utf8]{inputenc}
\usepackage[english]{babel}
\usepackage[nospace,noadjust]{cite}
\usepackage{fancyhdr}
\usepackage{lastpage}
\usepackage{enumerate}

\usepackage{cite}

\ifCLASSINFOpdf
  \usepackage[pdftex]{graphicx}
\else
  \usepackage[dvips]{graphicx}
\fi
\usepackage{graphicx}
\usepackage[caption=false,font=footnotesize]{subfig}
\usepackage{stackengine}

\usepackage{fixltx2e}

\usepackage{url}

\usepackage{textcomp}
\usepackage{bbm}
\usepackage{tikz}
\usetikzlibrary{ automata, arrows, positioning, calc, chains, shapes.multipart}

\IEEEoverridecommandlockouts

\ifCLASSINFOpdf
\else
\fi
\newcommand{\Ptx}{P^{on}_{tx}}
\newcommand{\Prx}{P^{on}_{rx}}
\newtheorem{proposition}{Proposition}
\newtheorem*{remark}{Remark}

\begin{document}

%
\title{Probabilistic Cooperation of a Full-Duplex Relay in Random Access Networks}

\author{\IEEEauthorblockN{Ioannis Avgouleas,  Nikolaos Pappas, Di Yuan, and Vangelis Angelakis}
	\IEEEauthorblockA{\\Department of Science and Technology,
				Link{\"o}ping University\\
				Norrk{\"o}ping SE-60174, Sweden\\
				Emails: \{ioannis.avgouleas, nikolaos.pappas,  di.yuan, vangelis.angelakis\}@liu.se}
	\thanks{ D. Yuan is also visiting professor at the Institute for Systems Research, University of Maryland, College Park, MD 20742, USA.}
	\thanks{The research leading to these results has received funding from the People Programme (Marie Curie Actions) of the European Union's Seventh Framework Programme FP7/2007-2013/ under REA grant agreements no [324515] (MESH-WISE) and no [612361] (SOrBet).}
}

\maketitle
\IEEEpeerreviewmaketitle

\begin{abstract}
In this work, we analyze the probabilistic cooperation of a full-duplex relay in a multiuser random-access network. 
The relay is equipped with on/off modes for the receiver and the transmitter independently. 
These modes are modeled as probabilities by which the receiver and the transmitter are activated.
We provide analytical expressions for the performance of the relay queue, such as arrival and service rates, stability conditions, and the average queue size.
We optimize the relay's operation setup to maximize the network-wide throughput while, simultaneously, we keep the relay's queue stable and minimize the consumed energy. 
Furthermore, we study the effect of the SINR threshold and the self-interference (SI) coefficient on the per-user and network-wide throughput. 
For low SINR threshold, we show under which circumstances it is beneficial to switch off the relay completely, or switch off the relay's receiver only.
\end{abstract}

\section{Introduction}
Internet of Things (IoT) devices' exponential proliferation and heterogeneous nature necessitate the development of ingenious methods for their communication and successful deployment. Due to the massive number of IoT devices and the insufficiency of available resources (e.g., energy, processing power etc.), several approaches have been presented for improvements of various aspects for their communication and interoperability. 

In this paper, we consider a wireless network with users trying to communicate with one destination node through a random access channel. We assume that users' communications are assisted by one intermediate node, the relay node. This cooperative communication setting can contribute towards the coverage, reliability and improvement of several Quality of Service (QoS) metrics of a wireless network \cite{Kramer-Cooperative-Communications-2006}. 

Among the cooperative techniques to increase QoS, full-duplex relaying, which we study in this paper, has proved itself to be a promising solution. Network-level cooperation can drastically increase throughput and reduce delay at the same time \cite{Rong-WiOpt2009,Rong-ISIT-2009}. It is not always beneficial in terms of the offered throughput and delay per packet to activate the relay  \cite{Pappas_TWC_2015}.
Therefore, to take the network's energy efficiency into account, we analyze the rate by which the relay node should be activated. 
We concentrate on the effect of a full-duplex cooperative relay, with on/off capabilities for the receiver and the transmitter independently, on several network performance metrics: throughput, arrival and service rates, as well as stability conditions.

\subsection{Related Work}
Within an IoT network, D2D communication can be used to assist users with exchanging information, and facilitate cooperation between user nodes and machine-type devices (MTDs) \cite{6231296}. D2D connections can also facilitate data aggregation, or relay information with the intention to reduce the number of connections between MTDs and base stations \cite{7248779}. 
For instance, D2D communications have been deployed as an underlay to long term evolution-advanced (LTE-A) networks \cite{IoT-Journal-2015-nrg-efficiency-D2D-res-alloc}. 
The proposed architecture exploits a cloud radio access network (C-RAN) to optimize the energy efficiency of each user with the assistance of C-RAN's distributed remote radio heads (RRHs), while a centralized interference mitigation algorithm, which runs in C-RAN's centralized baseband unit (BBU), improves the QoS performance.

Furthermore, many research activities have been devoted to making IoT more energy efficient. In \cite{Systems-journal-2015-enabling-tech}, the authors address the question of how the IoT enabling technologies can be efficiently orchestrated to achieve a green IoT network. Towards that end, the application of the time-reversal (TR) signal processing technique in the IoT context has been published in \cite{IoT-Journal-2014-time-reversal-green}. The authors argue that one of the most prominent characteristics of a TR system is that no coordination among separately located users is required. 
This leads to simplified MAC layer design, since there is no need for base stations in TR systems to perform coordination.

Relays have been introduced to improve several network performance metrics, such as coverage, outage probability, power efficiency and throughput \cite{relay-benefits-2004}. 
Nonetheless, most works study Half-Duplex (HD) relays due to their implementation simplicity, even though they do not utilize spectrum efficiently.
On the other hand, Full-Duplex (FD) relays, where nodes transmit and receive simultaneously, have recently attracted considerable attention both from industry and academia due to their potential to combat this problem.
Even though FD relays were considered impractical in the past, recent advancements combining different self-interference (SI) mitigation techniques, such as Propagation-Domain SI suppression, Analog-Circuit-Domain SI cancellation, and Digital-Domain SI mitigation, enable FD relays to offer the merits of both in-band FD wireless communications (high spectrum efficiency) and relaying (increased throughput, coverage, reliability, and decreased delay) \cite{ FD-Relay-Survey-2015, 5G-Full-Duplex-Apps-2015,5G-Survey-2015,5G-Self-Interference-Cancellation-2014}. 

Previous research efforts towards dynamically activating relays have been presented in the broad field of communications. 
For instance, \cite{OnDynamicPolicies-Friderikos-ICC2012} examines the potential of switching off wireless relays during low-utilization time periods to reduce the energy consumption of the network comparing to an \textquotedblleft always-on\textquotedblright ~policy. 
In \cite{Optimal-Relay-ICC-2011}, the authors optimize the relays' placement jointly with the relays on/off behavior to minimize the total power consumption, which is comprised of the transmission and the circuit power.

Network- or protocol-level cooperation has gained popularity in comparison to non-cooperative systems or physical-layer cooperation due to its potential to drastically increase throughput and significantly reduce the delay for all users \cite{Rong-WiOpt2009,Rong-ISIT-2009}. 
Cooperation improves performance gains as the channel quality improves \cite{Rong-WiOpt2009}. Seminal analyses and performance evaluations of network-level cooperation have been performed see e.g., \cite{Sadek-2009,Rong-ISIT-2009,Simeone-TCOMM-2007,Pappas-Globecom-2010,Pappas-ITW-2013}. 
The benefits of relaying depend on the network topology and nodes configuration such as channel power, packet arrivals etc. \cite{Simeone-TCOMM-2007} as well as the level of cooperation between the relay and the transmitting nodes \cite{Pappas-ISIT-2012}. 
Therein, the key idea is that a relay that is partially cooperating proves to perform equally or better to a fully cooperative relay. This idea gave rise to the concept we are studying in this paper, i.e., how to maximize performance in terms of network-wide throughput while keeping the relay node inactive as much as possible to reap the energy efficiency benefits.

\subsection{Contribution}
We extend and enhance the framework of \cite{Pappas_TWC_2015} to handle the case of a full-duplex relay node, which can switch on or off its transmitter and its receiver independently from each other.  
We provide analytical expressions for performance characteristics of the relay queue, such as the average arrival and service rates, as well as the average queue size. 
Also, we derive conditions for stability of the relay's queue as functions of the probability of activating the relay's receiver and the transmitter independently, the transmission probabilities, the self-interference coefficient, and the links’ outage probabilities.
Furthermore, we study the impact of the relay node and its activation probabilities on the per-user and the network-wide throughput.

The structure of this paper is organized as follows. 
Firstly, the system model along with the performance characteristics of the relay queue are described in Section \ref{sec:system-model}.
Then, in Section \ref{sec:relay-queue-performance}, two illustrative cases which help us gain significant intuition into the properties of the on/off FD relay in terms of the queue's probability of being empty, and its average size, are given.
Additionally in Section \ref{sec:throughput}, analytical expressions for the per-user and the network-wide throughput are given for the two cases.
The mathematical formulation for the relay's receiver and transmitter probabilities that maximize throughput while maintaining queue's stability are given in Section \ref{sec:optimization}.
Finally, numerical results are presented in Section \ref{sec:numerical-results}, and our conclusions are given Section \ref{sec:conclusion}.

\section{System Model}\label{sec:system-model}

\begin{figure}[t]
	\centering
	\includegraphics[width=1\columnwidth]{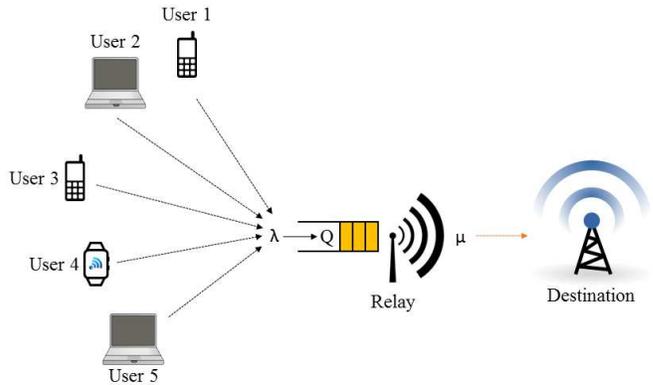}
 	\caption{Five users are transmitting their packets to the destination node. The full-duplex relay node assists users' transmissions by keeping the failed packets in its buffer (modeled as a queue $Q$ with arrival rate $\lambda$ and service rate $\mu$) and trying to transmit them in a later time-slot. 
	Users' direct links to the destinations are not depicted to facilitate readability.}
 	\label{fig:net}
\end{figure}

\subsection{Network Layer Model}
We consider a network with $n$ users, one full-duplex relay node with on/off capabilities for the receiver and the transmitter, independent of each other, and one destination node $d$. The relay's receiver and transmitter are not always on, thereby contributing to the overall energy efficiency. The receiver is on with probability $\Prx$, and the transmitter with probability $\Ptx$.
Users transmit packets to the destination with the cooperation of the relay node. 

We assume slotted time and that a packet transmission takes exactly one time-slot. Acknowledgments of successful transmissions are assumed instantaneous and error-free.
Furthermore, we assume multiple packet reception (MPR) for the relay and the destination node; i.e., more than one transmitter can successfully send packets to the same destination in the same time-slot.
If a user's transmission to the destination fails, the relay stores the missed packet into its queue and attempts to forward it to the destination later. 

Users are assumed to have random access to the medium, with no coordination among them, to represent scenarios where massive access to the channel is requested, and, as a result, centralized coordination of transmissions is infeasible or non-applicable.
We assume that users' queues are saturated i.e., have unlimited buffer size. The relay does not generate packet of its own; its purpose is to forward the users' packets. 
The relay attempts transmissions to the random access channel with probability $q_0$ when it has packets in its queue and its transmitter is on. The $i$-th user attempts transmissions with probability $q_i, \forall i =\{1,\dotsc,n\}$.                                   
                      
The wireless links are modeled as Rayleigh-fading ones with additive white Gaussian noise. A user's transmission is successful if the Signal-to-Interference-plus-Noise Ratio (SINR) between the receiver and the transmitter exceeds a threshold $\gamma$.
An instance of a topology of a simulated system with five users assisted by one relay is given in Fig. \ref{fig:net}.

\subsection{Performance Characteristics of Relay's Queue}
For presentation purposes, we will assume that $q_i = q, \forall i \in \{1,\dotsc,n\}$.
The \textit{average service rate} depends on $\Ptx$:
\begin{equation} \label{eq:srvRate_N_users}
\mu(\Ptx) = q_0 \Ptx \sum\limits_{k=0}^{n} \binom{n}{k} q^k (1-q)^{n-k} P_{0d,k},
\end{equation}
where $q_0$ is the attempt probability of the relay given it has packets in its queue, and $q$ is the user's attempt probability. $ P_{0d,k} $ is the success probability between the relay and the destination when $k$ nodes transmit.

The \textit{average arrival rate $\lambda$} depends on both $\Prx$ and $\Ptx$:
\begin{equation} \label{eq:arrRate_N_users}
\lambda(\Prx, \Ptx) = P(Q=0)\lambda_0 + P(Q>0)\lambda_1,
\end{equation}
where $\lambda_0 = \sum_{k=1}^{n} k r^0_k $ is the average arrival rate at the relay queue when the queue is empty, and $\lambda_1 = \sum_{k=1}^{n} k r^1_k  $ otherwise. The probability that the relay received $k$ packets when the queue is empty is denoted by $r^0_k$, or $r^1_k$ otherwise. 
The above mentioned performance characteristics depend on the success probabilities of the links between nodes, which are captured by the physical layer and are given in the following subsection.

\subsection{Physical Layer Model}
The wireless channel is modeled as Rayleigh flat-fading channel with additive white Gaussian noise. A packet transmitted by $i$ is successfully received by $j$ if and only if $ SINR(i,j) \geq \gamma_j$, where $\gamma_j$ is the threshold regarding the transmission to node $j$. 
Let $P_{tx}(i)$ be the transmit power of node $i$ and $r(i,j)$ be the distance between $i$ and $j$. Then, the power received by $j$ when $i$ transmits is $P_{rx}(i,j) = A(i,j)h(i,j)$, where $A(i,j)$ is a unit-mean exponentially distributed random variable representing channel fading. The receiver power factor $h(i,j)$ is given by $h(i,j) = P_{tx}(i)(r(i,j))^{- \alpha}$, where $\alpha \in [2,7]$ is the path loss exponent. 

Self-interference is modeled using the self-interference coefficient $g \in [0,1]$. The value of $g$ captures the accuracy of self-interference cancellation. When $g=1$, no self-interference cancellation technique is used, and when $g=0$ perfect self-interference is assumed. 
As $g$ approaches zero, the relay gets closer to pure full-duplex performance, while when $g$ approaches one, the relay tends to function more as a half-duplex one \cite{WeeraddanaWiOpt2010,WeeraddanaITW2010}.
The success probability in link $(ij)$ is given by \cite{Tse-book-2005}:

\begin{equation}
	\begin{aligned}\label{succ_prob}
		P^{j}_{i/\mathcal{T}} = exp\left(-\frac{\gamma_j \eta_j}{v(i,j)h(i,j)}\right) (1+\gamma_j r(i,j)^a g)^{-m} \times \\ \times \prod_{k \in \mathcal{T} \char`\\ \{i,j\} } \left( 1+\gamma_j \frac{v(k,j) h(k,j)}{v(i,j) h(i,j)} \right)^{-1},
	\end{aligned}
\end{equation}
where $\mathcal{T}$ is the set of transmitting nodes at the same time, $v(i,j)$ is the parameter of the Rayleigh fading random variable, $\eta_j$ is the receiving power at $j$, and $m=1$ when $j \in \mathcal{T}$ and $m=0$ elsewhere.

\section{Relay-Queue Performance}\label{sec:relay-queue-performance}
In the previous section, we laid the foundation to study the potential of the relay node on the per-user and network-wide throughput. We need first to define the probability the relay is empty, and the average queue size, though. The analysis is provided for two cases: (i) two asymmetric users, or (ii) an arbitrary number of symmetric users.

\subsection{Two Asymmetric Users}\label{sec:two-users}

The average arrival rate at the relay when its queue is empty is denoted by $\lambda_0$, or $\lambda_1$ otherwise. 
The relay queue size is denoted by $Q$. Let $\Prx$ and $\Ptx$ be the probability of the relay being switched on for reception and transmission respectively.  

The probability that the relay queue increases by $i$ packets when it is empty is denoted by $p^0_i$, or $p^1_i$ otherwise. 
Likewise, the probability that the queue decreases by one packet is $p^{1}_{-1}$.
The probability that the relay receives $i$ packets is denoted by $r^0_i$ when its queue is empty, and $r^1_i$ otherwise. 
In general $p^i_j \neq r^i_j$, since full-duplex relays allow simultaneous receptions and transmission, and, as a result, the probability of $i$ packets arriving is generally different to the probability of queue's increasing by $i$ packets.

For the sake of completeness and to facilitate the readability of the text, we quote the equations from \cite{Pappas_TWC_2015} along with the necessary extensions for the problem under consideration.
\begin{proposition}\label{thm:two-users}
The performance measures for the queue of a relay with on/off capabilities for the receiver and the transmitter in a two-users wireless network are the following:

	\begin{enumerate}[(i)] 
	\item The probability that the relay's queue is empty is given by: 
		\begin{equation}
		\begin{aligned}
		P(Q=0) = \frac{p^{1}_{-1} - p^{1}_{1} - 2p^{1}_{2}}{p^{1}_{-1} - p^{1}_{1} - 2p^{1}_{2} + \lambda_0 }.
		\end{aligned}
	\end{equation}
	\item The average queue size is given by:
\end{enumerate}
\begin{equation}
\begin{aligned} \label{eq:avgQueueSz_2users}
\bar{Q} &=  \frac{ (4p^0_1 + 10p^0_2) } {2(p^1_{-1} - p^1_1 - 2p^1_2 + \lambda_0 )} +\\
&+\frac{\lambda_0 (2p^1_{-1}-4p^1_1-10p^1_2)}{2(p^1_1 + 2 p^1_2 - p^1_{-1})(p^1_{-1} - p^1_1 - 2p^1_2 + \lambda_0 )}.
\end{aligned}
\end{equation}
\end{proposition}

\begin{IEEEproof} See Appendix \ref{appxA}. \end{IEEEproof}

\begin{remark}
The calculations of the average arrival rate, when the queue is empty, $\lambda_0$, the average arrival rate, when the queue is not empty, $\lambda_1$, and the value of $q_0$ for which the relay's queue is stable are given in Appendix \ref{appxA} as well. Queue stability is important, since having bounded queue size corresponds to finite queuing delay and guarantees that all packets will be transmitted.
\end{remark}

\subsection{Symmetric Users}\label{nSymmetricUsers}
We include the case of $n>2$ users which are distributed symmetrically around the relay and attempt transmissions with the same probability. Such consideration is a good representation of real IoT networks with a huge amount of users, and, hence, we can gain insight into how several performance measures scale up in the number of users.

The average arrival rate is denoted by $\lambda_0$ or $\lambda_1$, when the queue is empty or not respectively. The former is a function of $\Prx$ and the latter is a function of both $\Prx$ and $\Ptx$.
The probability that the relay's queue increases by $i$ packets when its queue is empty, or not, are denoted by $p^0_i$ and $p^1_i$ respectively, similarly to the two-user case.
Both $p^0_i$ and $p^1_i$ depend on $\Prx$ and $\Ptx$.
The probability that the queue decreases by one packet is $p^{1}_{-1}$ and depends only on $\Ptx$.

Similarly to Section \ref{sec:two-users}, we quote the equations from \cite{Pappas_TWC_2015} along with the necessary adaptations to contribute to the coherence of the presentation.
\begin{proposition}\label{thm:n-users}
The performance measures for the queue of one relay with on/off capabilities for the receiver and the transmitter in a wireless network with n-symmetric users are the following:

\begin{enumerate}[(i)] 
	\item The probability that the relay's queue is empty depends on both $\Prx$ and $\Ptx$:
	\begin{equation}\label{eq:probRelayEmpty_Nusers}
	\begin{aligned}
	P(Q=0) = \frac{p^{1}_{-1} - \sum\limits_{i=1}^{n} i  p^1_i }{p^{1}_{-1} - \sum\limits_{i=1}^{n} i  p^1_i + \lambda_0 }.
	\end{aligned}
	\end{equation}

	\item The average queue size depends on both $\Prx$ and $\Ptx$:
\end{enumerate}
\begin{equation} \label{eq:avgQueueSz_Nusers} 
\begin{aligned}
\bar{Q} &=\frac{ (  \sum\limits_{i=1}^{n} i p^1_i - p^1_{-1} ) \sum\limits_{i=1}^{n}i(i+3)p^0_i} {2 ( p^1_{-1} - \sum\limits_{i=1}^{n} i p^1_i + \lambda_0 ) } + \\
&+\frac{\lambda_0 (2p^1_{-1}- \sum\limits_{i=1}^{n}i(i+3)p^1_i ) }{ 2 (\sum\limits_{i=1}^{n} i p^1_i - p^1_{-1} ) ( p^1_{-1} - \sum\limits_{i=1}^{n} i p^1_i + \lambda_0 ) }.
\end{aligned}
\end{equation}
\end{proposition}

\begin{IEEEproof}
See Appendix \ref{appxB}.
\end{IEEEproof}
\begin{remark}
The values of $q_0$ for which the relay's queue is stable are given by $q_{0min} < q_0 < 1$, where $q_{0min}$ is derived in Appendix \ref{appxB} as well.
\end{remark}

\section{Maximizing Throughput and Maintaining Queue Stability}\label{sec:throughput}

The \textit{per-user throughput}, $T_i$, for the $i$-th user is given by:
\begin{equation} \begin{aligned}
	T_i = T_{D,i} + T_{R,i},
\end{aligned} \end{equation}
where $T_{D,i}$ denotes the \textit{direct throughput} from user $i$ to the destination $d$ i.e., without the relay's assistance.
When the transmission at the destination is not successful, and the relay node has correctly received the packet, then it stores it to its queue. The contributed throughput by the relay to the user $i$ is denoted by $T_{R,i}$.
The \textit{network-wide throughput} is given by: 
$ T_{net} = \sum_{i=1}^{n} T_i $.

\subsection{Per-User and Network-wide Throughput: Two users}
The \textit{direct throughput} to the destination for the $i$-th user depends on both $\Prx$ and $\Ptx$, and is given by: 

\begin{multline}
T_{D,i}(\Prx, \Ptx) = \\
q_0 \Ptx P(Q>0) q_i \Big[ (1-q_j) P^{d}_{i/0,i} + q_jP^{d}_{i/0,i,j} \Big] + \\
+~ [1-q_0 \Ptx P(Q>0)]	q_i \Big[ (1-q_j) P^{d}_{i/i} + q_jP^{d}_{i/i,j} \Big].
\end{multline}
Recall that the relay and the $i$-th user attempt transmissions with probabilities $q_0$ and $q_i$ respectively.

The \textit{relayed throughput} $T_{R,i}$ of user $i$ is given by:
\begin{align}
&	T_{R,i}(\Prx, \Ptx) = q_0 \Ptx P(Q>0) q_i \times \nonumber\\
&	\times \Big[ (1-q_j) (1-P^{d}_{i/0,i})P^{0}_{i/0,i} + q_j (1-P^{d}_{i/0,i,j}) P^{0}_{i/0,i,j} \Big] + \nonumber\\
&	+ [1-q_0 \Ptx P(Q>0)] q_i \Big[ (1-q_j) (1-P^{d}_{i/i})P^{0}_{i/i} + \nonumber\\
&   + q_j (1-P^{d}_{i/i,j})P^{0}_{i/i,j} \Big].
\end{align}

\subsection{Per-User and Network-wide Throughput: $n$ Symmetric Users}

We denote the \textit{per-user throughput} by $T$, the \textit{direct throughput} to the destination by $T_D$, and the \textit{relayed throughput} by $T_R$. The relay and every (symmetric) user attempt transmissions with probabilities $q_0$ and $q$ respectively.

The direct throughput $T_D$ depends on both $\Prx$ and $\Ptx$, and is given by:
\begin{multline} \label{eq:Td_N_users}
T_{D}(\Prx, \Ptx) = q_0 \Ptx P(Q>0) \times \\
\times \sum\limits_{k=0}^{n-1} \binom{n-1}{k} q^{k+1} (1-q)^{n-k-1} P_{d,k+1,1} + \\
+~ [1-q_0 \Ptx P(Q>0)]
\sum\limits_{k=0}^{n-1} \binom{n-1}{k} q^{k+1} (1-q)^{n-k-1} P_{d,k+1,0}.
\end{multline}

The \textit{relayed throughput} $T_R$, with the assumption that the relay queue is stable (see Section \ref{nSymmetricUsers}), depends on both $\Prx$ and $\Ptx$, and is given by:\textsl{}
\begin{multline} \label{eq:Tr_N_users}
T_{R}(\Prx, \Ptx) = q_0 \Ptx P(Q>0) \times \\
\times \sum\limits_{k=0}^{n-1} {n-1 \choose k} q^{k+1} (1-q)^{n-k-1}(1-P_{d,k+1,1}) P_{0,k+1,1} \\
+ [1-q_0 \Ptx P(Q>0)]  \times \\
\times \sum\limits_{k=0}^{n-1} {n-1 \choose k} q^{k+1} (1-q)^{n-k-1} (1-P_{d,k+1,0}) P_{0,k+1,0},
\end{multline}
where $P_{d,i,j}$ and $P_{0,i,j}$ denote the success probabilities for the symmetric case given in Section \ref{nSymmetricUsers}.

Hence, the \textit{per-user throughput} $T$, when the relay queue is stable, is given by:
\begin{equation*}\label{throughput}
 T = T_D + T_R.
\end{equation*}
The \textit{network-wide throughput} for $n$ symmetric users is: 
\[ T_{net} = nT. \]
\subsection{Optimization Formulation}\label{sec:optimization}

We present the optimization problem for maximizing throughput, while guaranteeing the relay's queue stability, by adjusting the activation probabilities for the relay's receiver ($\Prx$) and transmitter ($\Ptx$). 
The general optimization formulation is given by:

\begin{equation}\label{optimization_params}
\left\{ \begin{array}{cc}
max. & T_{D}(\Prx, \Ptx) + T_{R}(\Prx, \Ptx) \\
s.t. & \lambda(\Prx, \Ptx ) < \mu(\Ptx)\\
& 0 \leq \Prx, \Ptx \leq 1\\
\end{array} \right\} 
\end{equation} 
There are additional constraints that are specific to if there is one or multiple users.
The following subsections analyze the optimization constraints for the two respective cases.

\subsubsection{\textbf{One user}}
In case there is one user, the direct and relayed throughput depend on $\Prx$ and $\Ptx$, similarly to (\ref{eq:Td_N_users}) and (\ref{eq:Tr_N_users}):
\begin{multline}
T_{D}(\Prx,\Ptx) = q_0 \Ptx P(Q>0) q_{1} P^{d}_{1/0,1} + \\
+ \Big[1-q_0 \Ptx P(Q>0)\Big] q_{1} P^{d}_{1/1},
\end{multline}
\begin{multline}
T_{R}(\Prx,\Ptx) = q_0 \Ptx P(Q>0) q_{1} \Big[ 1 - P^{d}_{1/0,1}  \Big]  P^{0}_{1/0,1} + \\
+\Big[ 1-q_0 \Ptx P(Q>0) \Big] q_{1} \Big[ 1 - P^{d}_{1/1}  \Big] P^{0}_{1/1}, 
\end{multline}
Additionally, the average arrival and service rate are respectively given by:
\begin{equation}
\begin{aligned}
\lambda &=& \frac{p^{1}_{-1}-p^{1}_{1}}{p^{1}_{-1}-p^{1}_{1}+p^{0}_{1}}\lambda_0 + \frac{p^{0}_{1}}{p^{1}_{-1}-p^{1}_{1}+p^{0}_{1}}\lambda_1, \\
\mu &=&  q_0 \Ptx \Big(q_1 P^{d}_{0/0,1}+(1-q_1)P^{d}_{0/0}\Big), 
\end{aligned}
\end{equation}
where $\lambda_0$ and $\lambda_1$ denote the average arrival rate, when the queue is empty or not respectively. The probability that the relay queue is increased by $i$ packets when it is empty, $p^0_i$, or not empty, $p^1_i$, and the probability that the queue is decreased by one packet are given by:

\begin{flalign*}
p^{0}_{1} &=  q_1 \Big(1-P^{d}_{1/1}\Big) P^{0}_{1/1},& \\
p^{1}_{1} &= (1 - q_0) q_1 \Big(1-P^{d}_{1/1}\Big) P^{0}_{1/1} +& \\
&+q_0 \Ptx q_1 \Big(1-P^{d}_{1/0,1}\Big) P^{0}_{1/0,1} (1-P^{d}_{0/0,1}),& \\
p^{1}_{-1} &= q_0 \Ptx (1-q_1) P^{d}_{0/1} + q_0 \Ptx q_1 P^{d}_{0/0,1} P^{d}_{1/0,1} +& \\
&+ q_0 \Ptx q_1 \Big(1-P^{d}_{1/0,1}\Big) \Big( 1-P^{0}_{1/0,1} \Big) P^{d}_{0/0,1},&
\end{flalign*}
where $P^{d}_{i/T}$ is the success probability between node $i$ and node $d$ while the transmitting nodes constitute set $T$. This probability can be calculating using (\ref{succ_prob}).

\subsubsection{\textbf{$n$ users}}
We consider the $n$-symmetric users case of Section \ref{nSymmetricUsers}
with the direct and relayed throughput given by $T_{D}(\Prx, \Ptx)$ and $T_{R}(\Prx, \Ptx)$ of (\ref{eq:Td_N_users}) and (\ref{eq:Tr_N_users}) respectively. The average service rate $\mu(\Ptx)$ and the average arrival rate $\lambda(\Prx, \Ptx)$ are given by (\ref{eq:srvRate_N_users}) and (\ref{eq:arrRate_N_users}) respectively. As a result, relay's queue stability  is satisfied due to: \[\lambda(\Prx, \Ptx) < \mu(\Ptx).\]

The optimization problem in (\ref{optimization_params}) is not a convex one, and, thus, it's hard to find a closed form solution. 
We evaluate the results numerically using a nonlinear solver. The results can be found in Section \ref{sec:numerical-results}.

\section{Numerical Results}\label{sec:numerical-results}
In this section, we provide numerical results to validate the aforementioned analysis. We consider the case where all users have the same link characteristics and transmission probabilities to facilitate exposition clarity. The specific parameters we used for our numerical results can be found in table \ref{parameters_table}.
\begin{table}[!t]
	\centering
	\renewcommand{\arraystretch}{1.3}
	\caption{Simulation parameters}
	\label{parameters_table}
	\begin{tabular}{|c|c|c|}
		\hline
		& \textbf{Explanation} & \textbf{Value}\\
				\hline
		$ r(0,d)$ & relay-destination distance	& $80$ m\\
				\hline
		$ r(i,0), \forall i \in \{1,\dotsc,n\}$ & user-relay distance	& $60$ m\\
				\hline
		$ r(i,d), \forall i \in \{1,\dotsc,n\}$ & user-destination distance & $130$ m\\				
				\hline								
		$ \alpha $ & link path loss exponent & $4$\\
				\hline
		$ P_{tx}(i), \forall i \in \{1,\dotsc,n\}$ & transmit power of the user & $1$ mW\\
				\hline
		$ P_{tx}(0) $ & transmit power of the relay & $10$ mW\\
				\hline
		$ q_i, \forall i \in \{1,\dotsc,n\}$ & user transmission attempt probability & $0.1$\\
				\hline
		$ n_0 $ & receiver noise power & $10^{-11}$\\
				\hline
		
	\end{tabular}
\end{table}

\subsection{Per-user and Network-wide Throughput}
\begin{figure*}[!htbp]
	\begin{center}
		\subfloat{
		\stackunder[5pt]{\includegraphics[width=1\columnwidth]{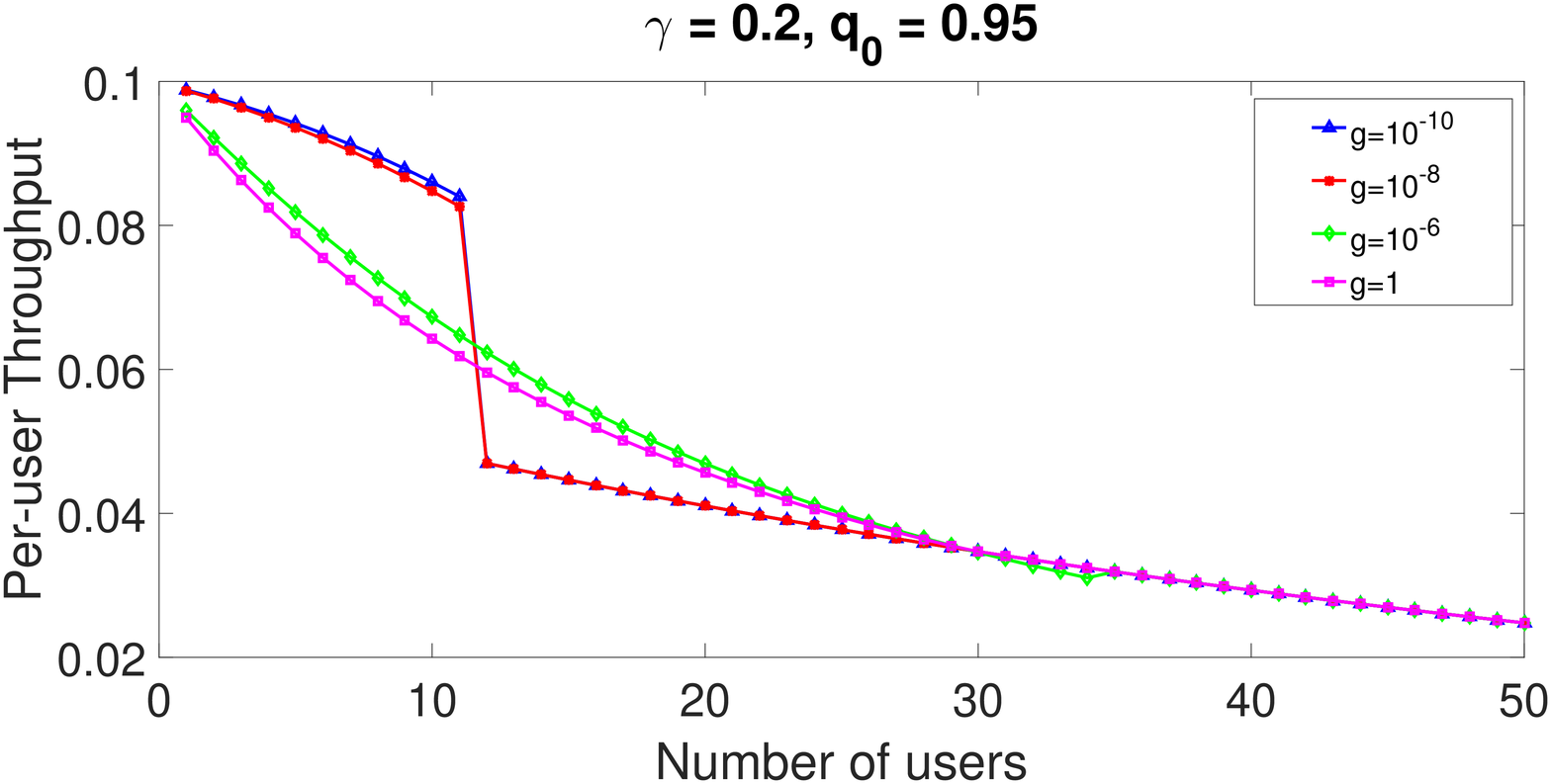}}{(a)}
		\stackunder[5pt]{\includegraphics[width=1\columnwidth]{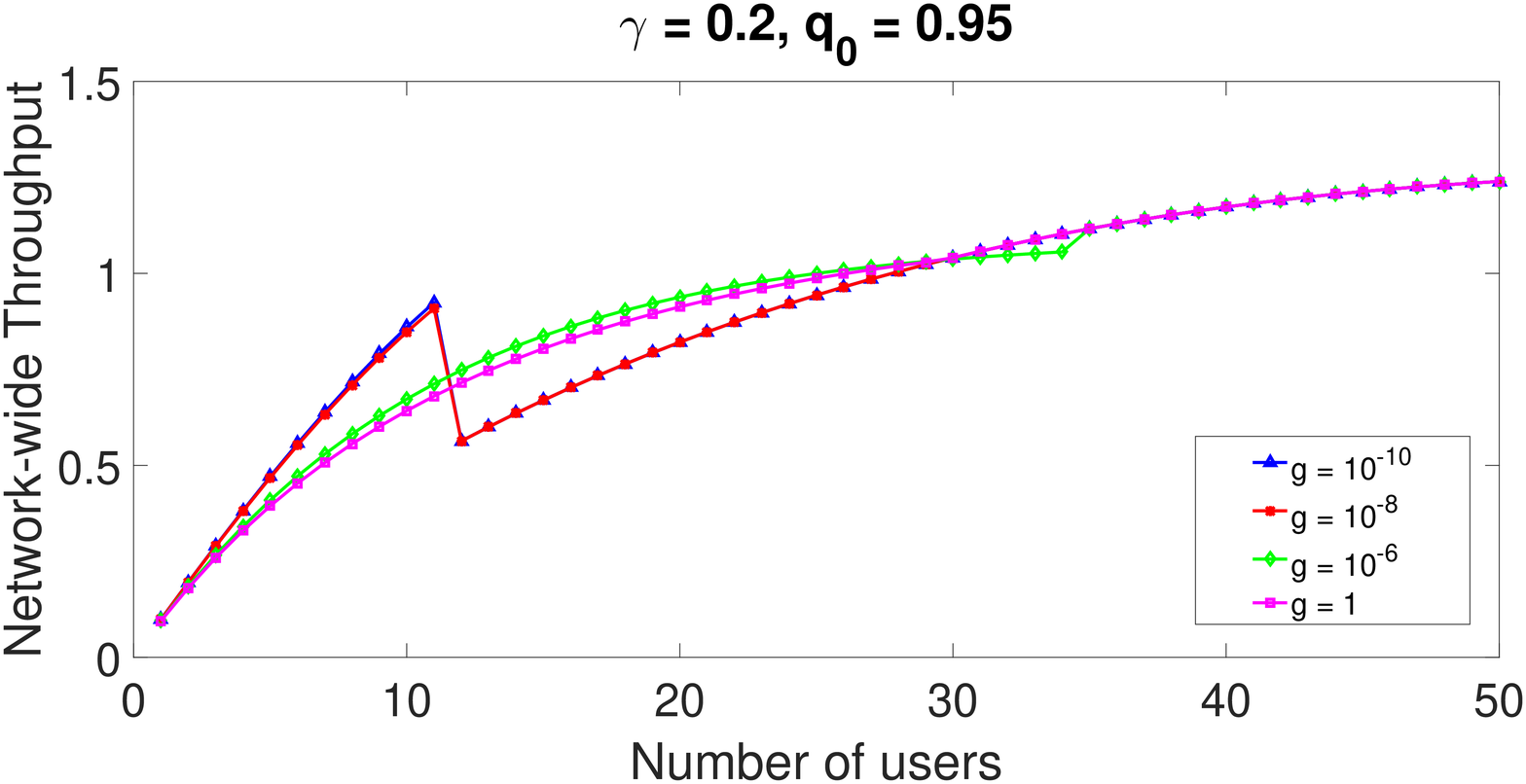}}{(b)}} 
	\end{center}	
	\caption{Per-user and network-wide throughput vs. the number of users for $\gamma=0.2, q=0.1$ and $q_0=0.95$. (a) Per-user throughput vs. the number of users. (b) Network-wide throughput vs. the number of users.}
	\label{fig:throughputs02}	
	\begin{center}
		\subfloat{ \stackunder[5pt]{\includegraphics[width=1\columnwidth]{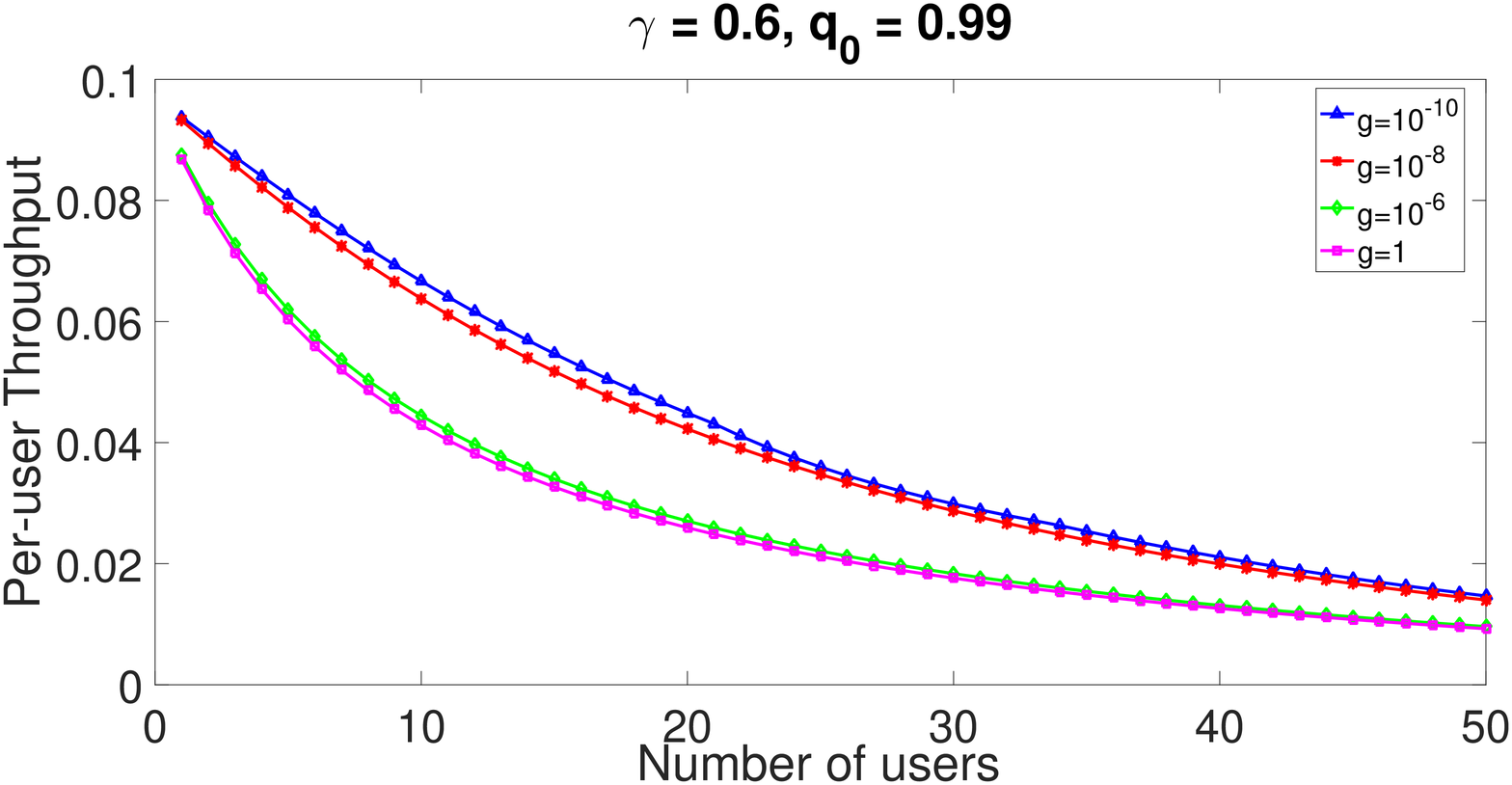}}{(a)} \stackunder[5pt]{\includegraphics[width=1\columnwidth]{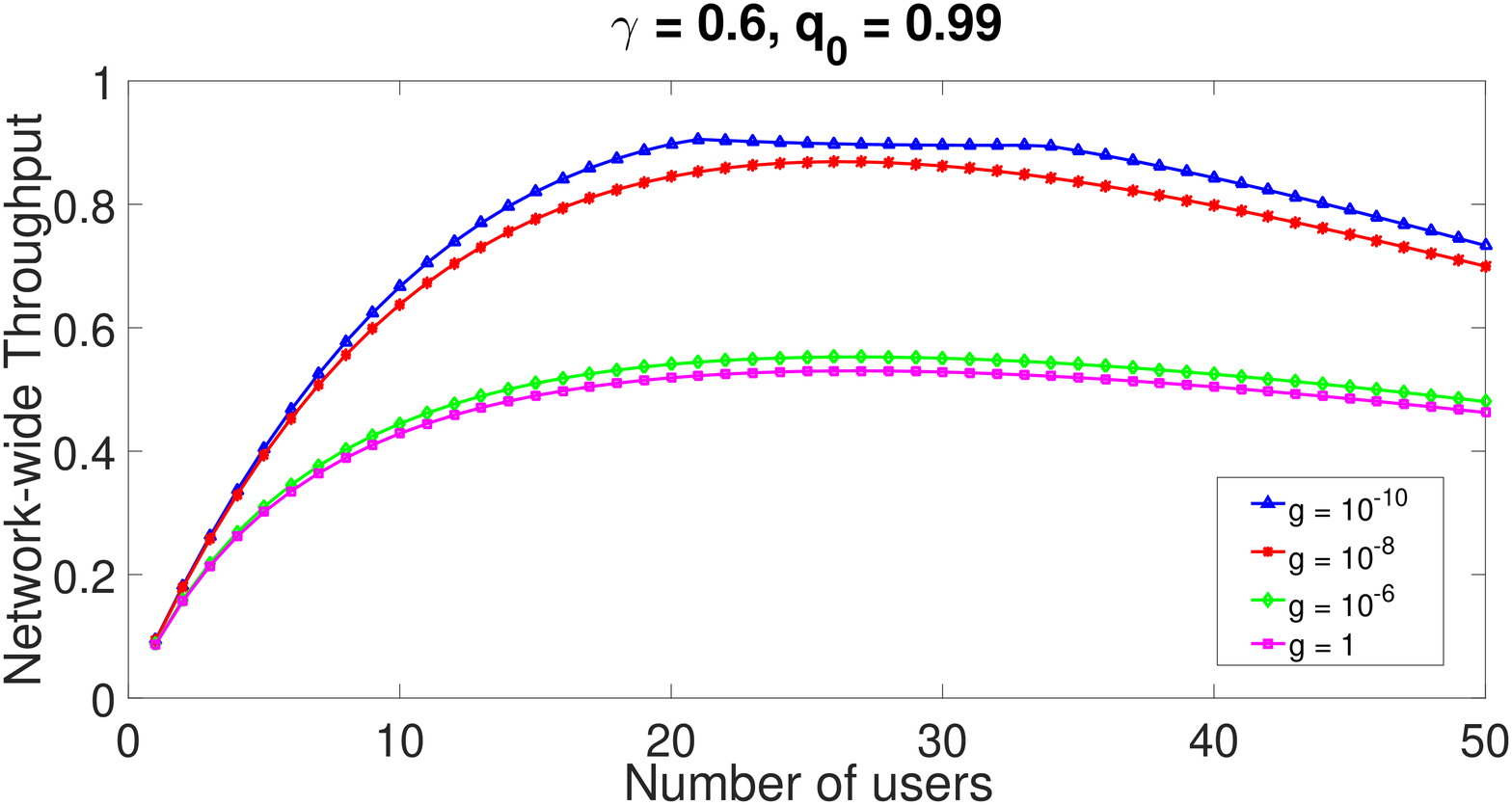}}{(b)} } 
	\end{center}	
	\caption{Per-user and network-wide throughput vs. the number of users for $\gamma=0.6, q=0.1$ and $q_0=0.99$. (a) Per-user throughput vs. the number of users. (b) Network-wide throughput vs. the number of users.}
	\label{fig:throughputs06}	
	\begin{center}
		\subfloat{ \stackunder[5pt]{\includegraphics[width=1\columnwidth]{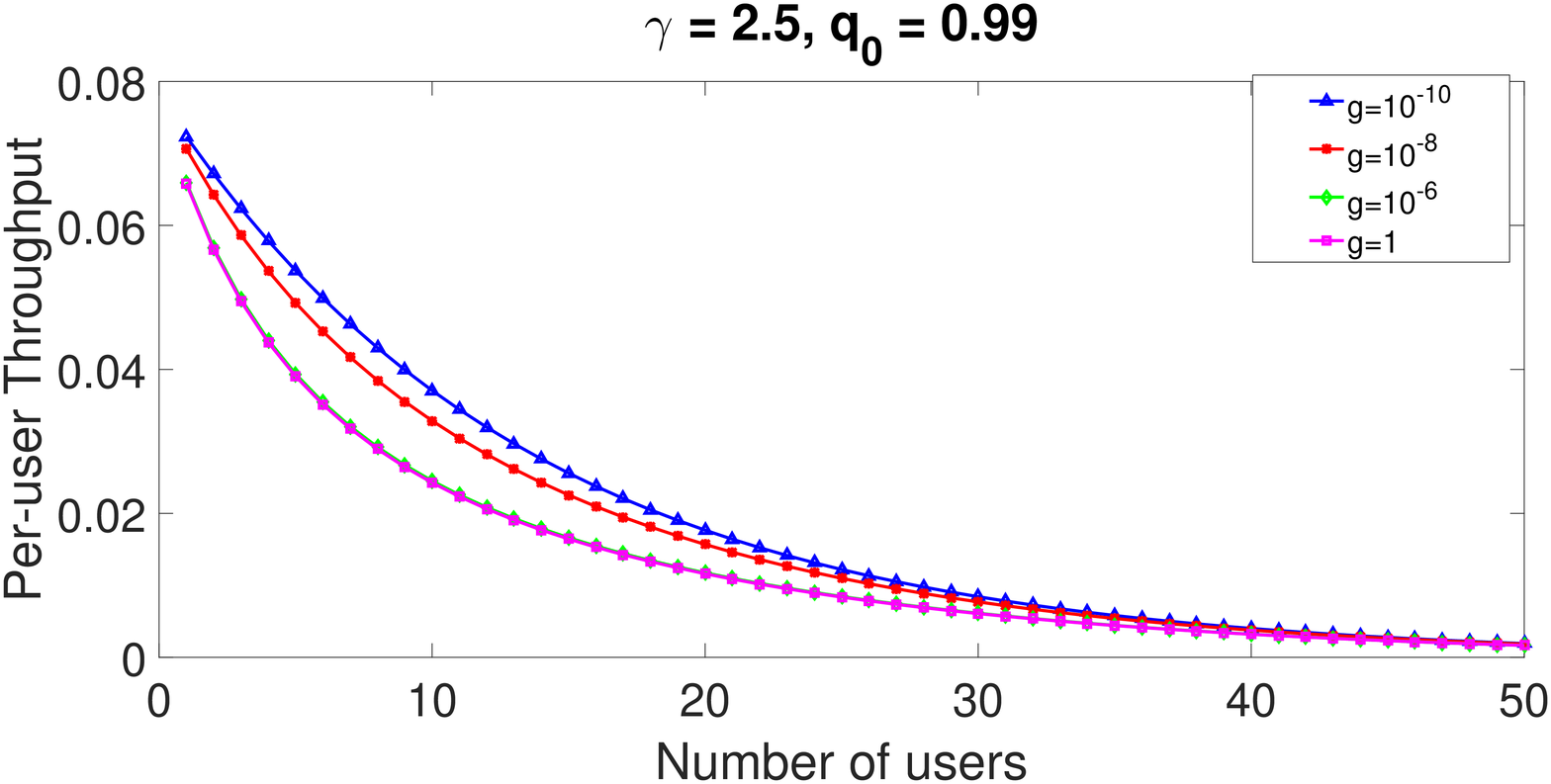}}{(a)}
		\stackunder[5pt]{\includegraphics[width=1\columnwidth]{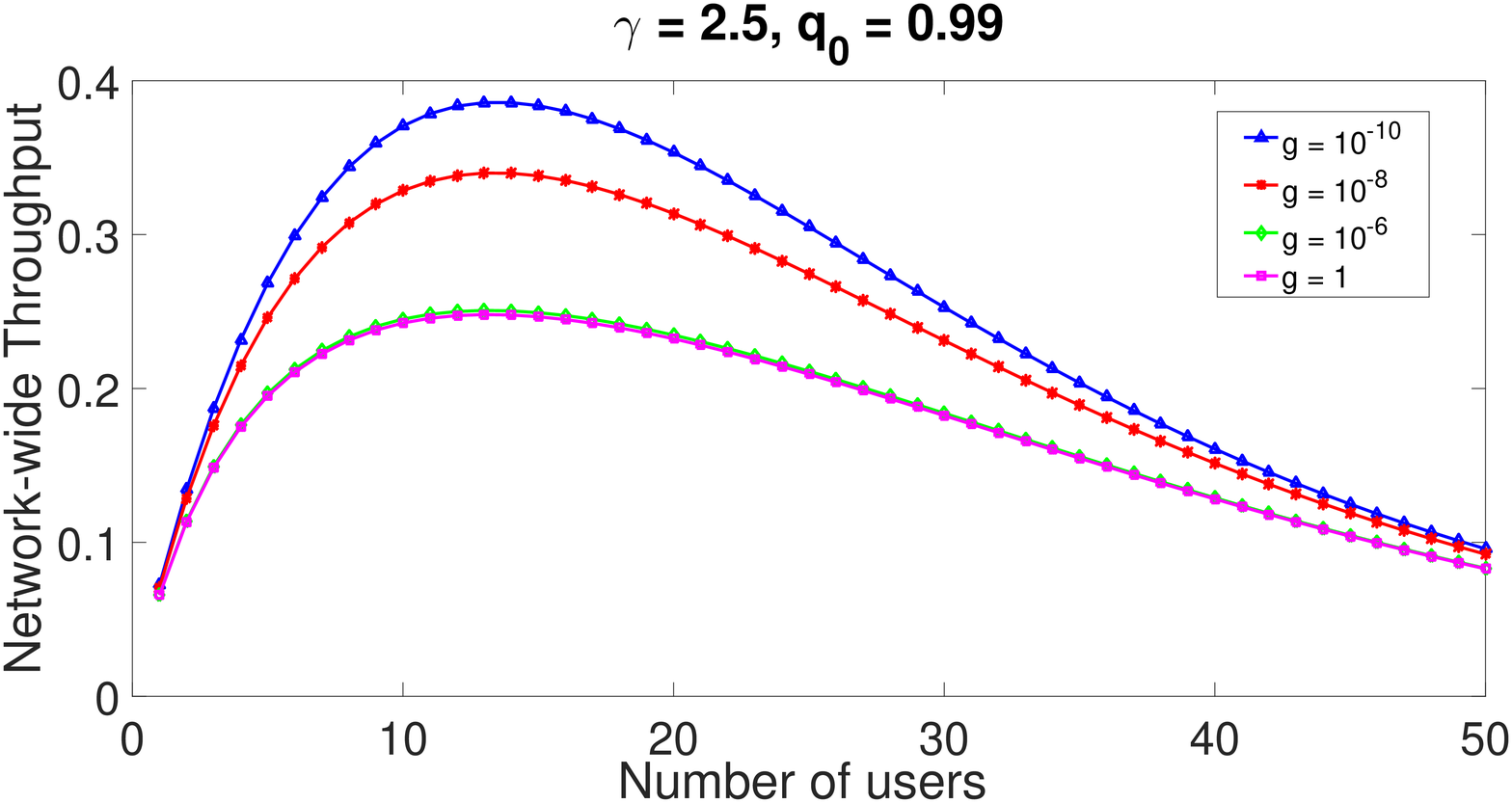}}{(b)} } 
	\end{center}	
	\caption{Per-user and network-wide throughput vs the number of users for $\gamma=2.5, q=0.1$ and $q_0=0.99$. (a) Per-user throughput vs. the number of users. (b) Network-wide throughput vs. the number of users.}
	\label{fig:throughputs25}	
\end{figure*}

Figs. \ref{fig:throughputs02}(a),  \ref{fig:throughputs06}(a), and \ref{fig:throughputs25}(a) present the per-user throughput versus the number of users using the optimized values for the activation probabilities of the relay's receiver ($\Prx$), and transmitter ($\Ptx$) on (see Section \ref{sec:optimization}). 
Figs. \ref{fig:throughputs02}(b),  \ref{fig:throughputs06}(b), and \ref{fig:throughputs25}(b) demonstrate the corresponding network-wide throughput versus the number of users.
We have used different values for the self-interference coefficient $g,q_0$ and $\gamma$ to gain insight into the performance measures under consideration. 

When $\gamma=0.2$, it is highly probable that there will be more simultaneous successful transmissions in the system comparing to higher $\gamma$ values (e.g., $0.6$ or $1.2$). As a result, the relay's queue is more probable to be unstable, since it can receive multiple packets, but transmit at most one, in every time-slot. 
The first constraint in (\ref{optimization_params}) forces the stability of the queue, which is achieved by appropriately decreasing the receiver's probability $\Prx$, increasing the transmitter's probability, or combining both methods.
It also justifies the drop in the per-user throughput for a relatively small number of users (e.g., less than $12$), when $g=10^{-10}$ and $g=10^{-8}$ (see Fig. \ref{fig:throughputs02}). 
In comparison to the results of \cite{Pappas_TWC_2015} where no traffic control on the relay's queue was considered, less throughput is experienced due to the stability of the queue along with the respective energy efficiency gains (since $\Prx$ and $\Ptx$ have to be less than one to keep the queue stable).

\begin{figure*}[!htbp]
	\begin{center}
		\subfloat{ 
			\stackunder[3pt]{\includegraphics[width=1\columnwidth]{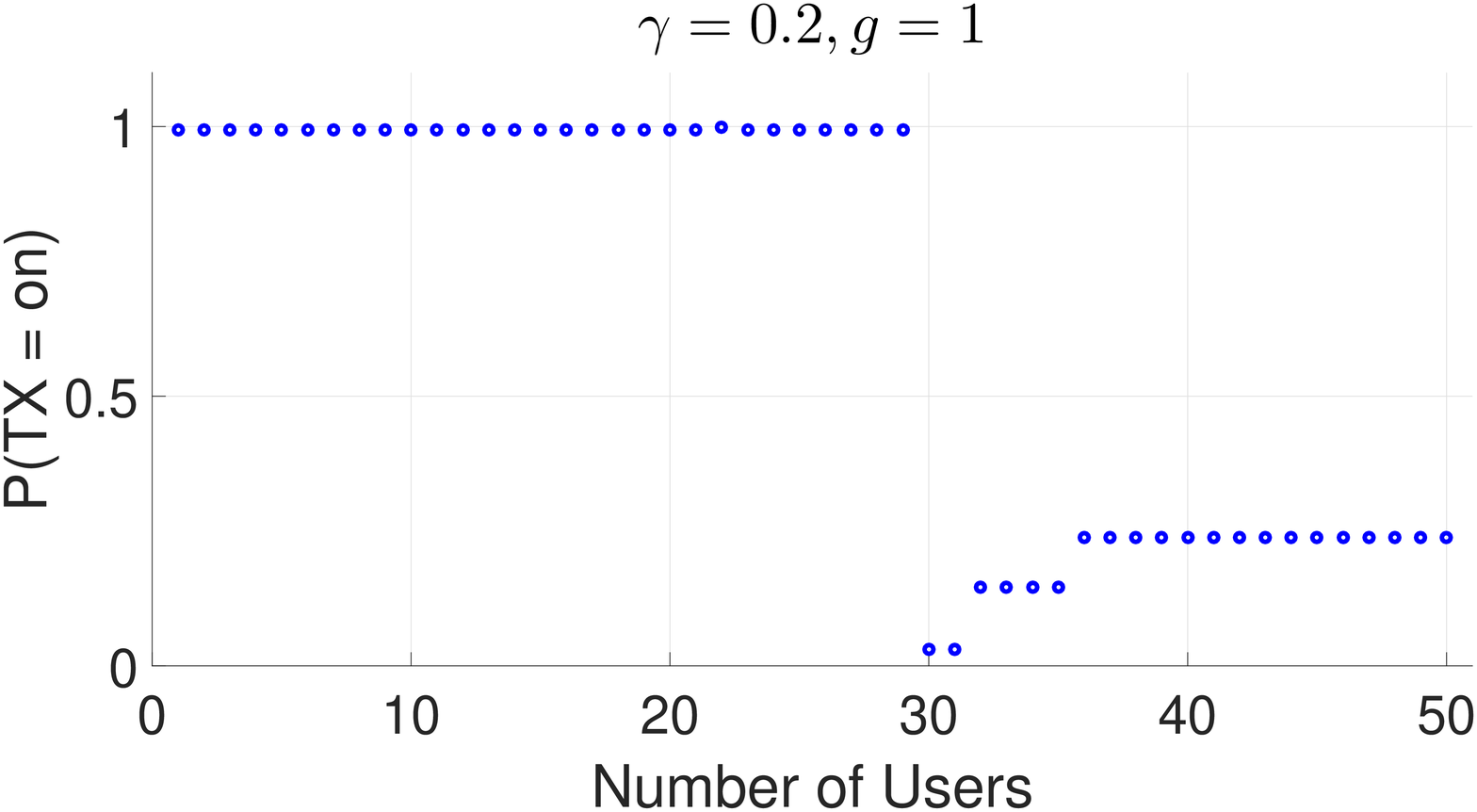}}{(a)}
			\stackunder[3pt]{\includegraphics[width=1\columnwidth]{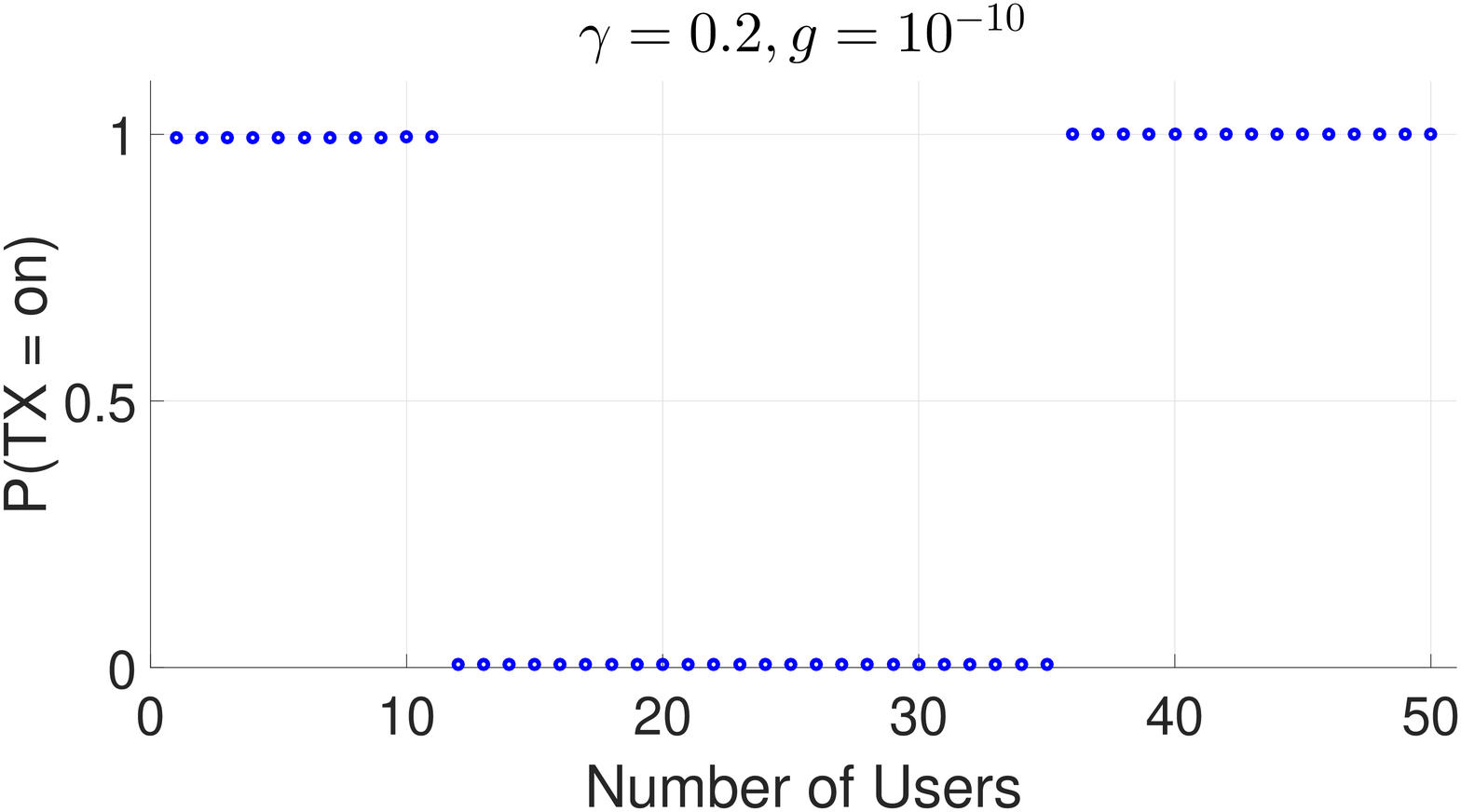}}{(b)} } 
	\end{center}	
	\caption{Probability that the relay's transmitter (TX) should be activated to maximize throughput vs the number of users for $\gamma=0.2, q=0.1$ and $q_0=0.95$, when (a) the self-interference cancellation coefficient $g=1$ (resembling HD relay operation), and (b) $g=10^{-10}$ (resembling FD relay operation).} 
	\label{fig:probTX_02}	
	\begin{center}
		\subfloat{ 
			\stackunder[3pt]{\includegraphics[width=1\columnwidth]{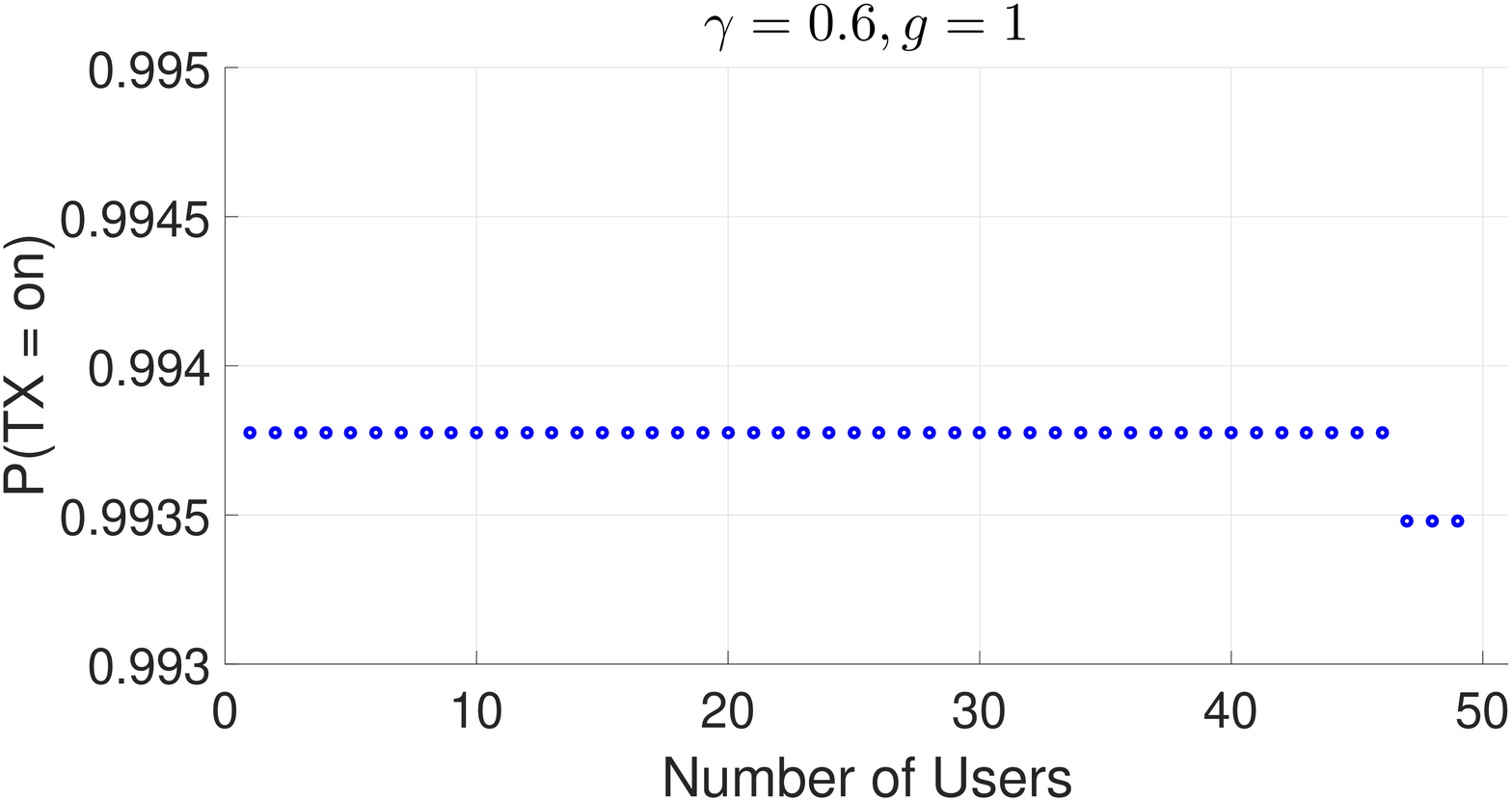}}{(a)}
			\stackunder[3pt]{\includegraphics[width=1\columnwidth]{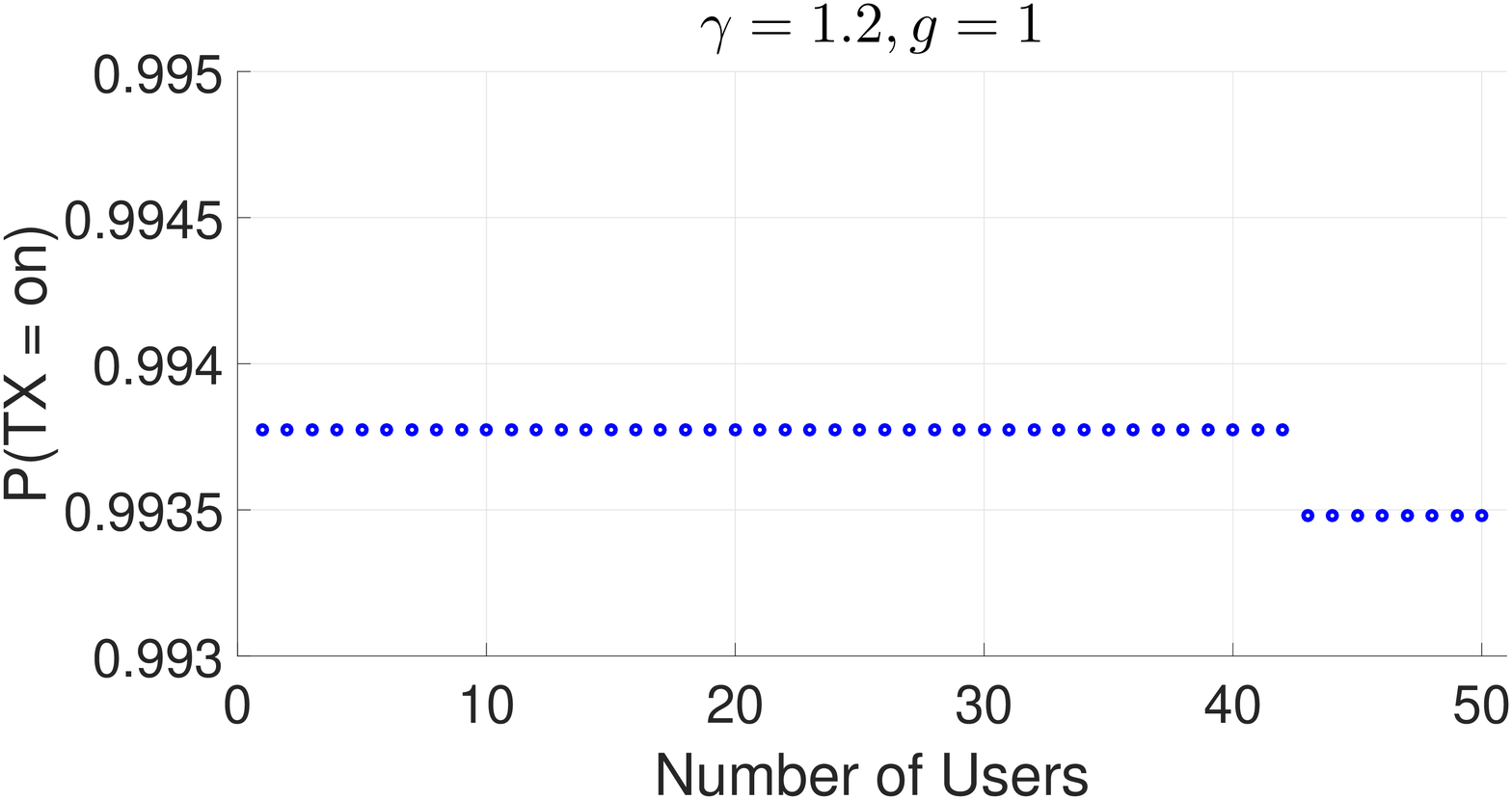}}{(b)} }
	\end{center}	
	\caption{Probability that the relay's transmitter (TX) should be activated to maximize throughput vs the number of users for $q=0.1, ~q_0=0.99$ and $g = 1$ (resembling FD relay operation), when (a) the SINR threshold $\gamma=0.6$, and (b) $\gamma=1.2$.}
	\label{fig:probTX_06plus12}
	\begin{center}
		\subfloat{ 
			\stackunder[3pt]{\includegraphics[width=1\columnwidth]{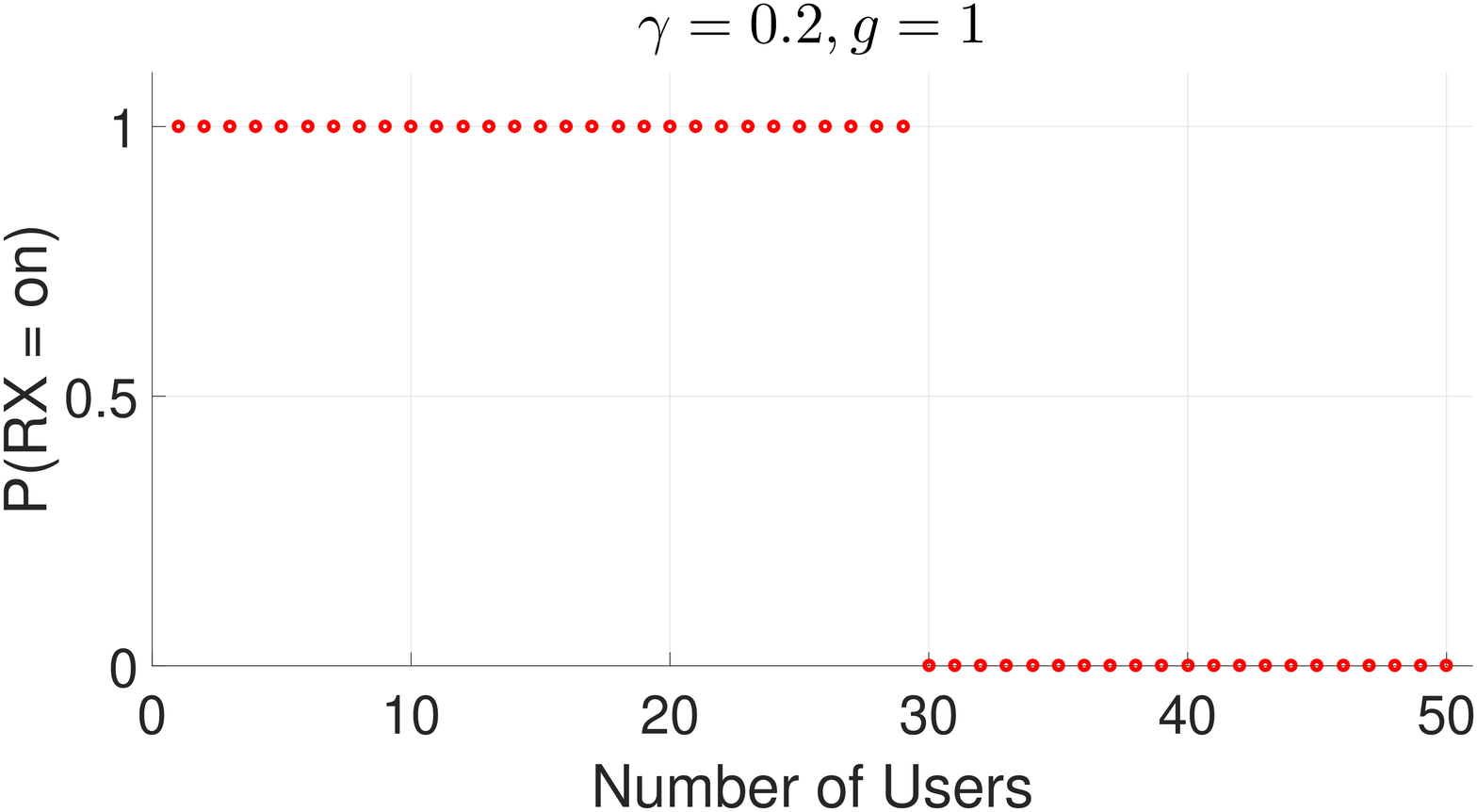}}{(a)}
			\stackunder[3pt]{\includegraphics[width=1\columnwidth]{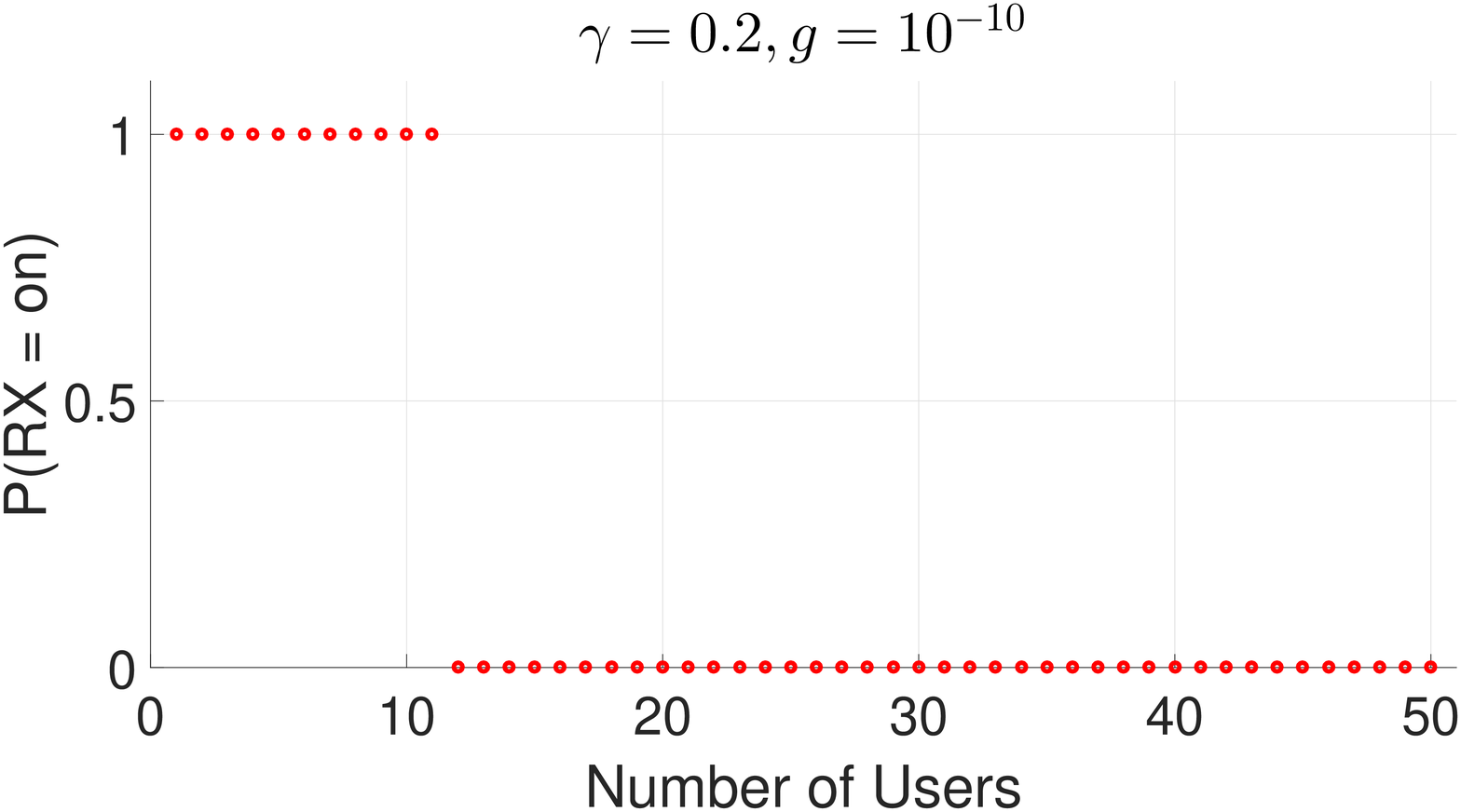}}{(b)} } 
	\end{center}	
	\caption{Probability that the relay's receiver (RX) should be activated to maximize throughput vs number of users for $\gamma=0.2, q=0.1$ and $q_0=0.99$, when (a) the self-interference cancellation coefficient $g=1$ (resembling HD relay operation), and (b) $g=10^{-10}$ (resembling FD relay operation).}
	\label{fig:probRX_02}
\end{figure*}

For $\gamma=0.6$, $\Prx$ and $\Ptx$ have to be decreased to keep the queue stable. However, this decrease is not substantial (see e.g, Fig. \ref{fig:probTX_06plus12}(a)) and, no considerable gains, compared to \textquotedblleft always-on\textquotedblright ~operation, in terms of energy efficiency are observed. 
In this case and when $\gamma=2.5$ (Figs. \ref{fig:throughputs06} and \ref{fig:throughputs25}), throughput gains are considerable using perfect SI cancellation (i.e., $g = 10^{-10}$ and $g = 10^{-8}$) compared to using no SI cancellation at all (i.e., $g = 10^{-6}$ and $g = 1$).

\subsection{Optimal Receiver and Transmitter Activation Probabilities.}

In this subsection, we present the optimal values of the activation probabilities of the relay's receiver, $\Prx$, and transmitter, $\Ptx$ that maximize throughput and, simultaneously, maintain the relay's queue stability (see Section \ref{sec:optimization} for the mathematical formulation).
Lower $\Prx$ implies lower consumed energy, and lower $\Ptx$ results in lower consumed energy as well as lower interference to the direct links of the users, since the relay transmits less frequently. The clear effect of the latter is explained by that, according to our parameters, the relay uses $10$ times higher power than a user.
The optimization of the activation probabilities allows for maximizing the offered throughput while simultaneously being as energy efficient as possible. 

Figs. \ref{fig:probTX_02} and \ref{fig:probTX_06plus12} present the values of $\Ptx$ that maximize throughput and maintain queue stability. 
When $g=1$ is used to resemble HD relay operation, always activating the relay's transmitter maximizes throughput, except if a relatively high number of users (e.g., over $30$) is simultaneously transmitting and the SINR threshold $\gamma=0.2$ (Fig. \ref{fig:probTX_02}). 
Deactivating almost completely the relay's transmitter is beneficial when serving a relatively small number of users (less than $30$) and the relay operates in FD mode (e.g., $g=10^{-10}$). 
On the other hand, when $\gamma$ takes higher values (e.g., $0.6$ or $1.2$) the transmitter should be kept nearly \textquotedblleft always-on\textquotedblright ~(Fig. \ref{fig:probTX_06plus12}), and, hence, more energy will be used to achieve maximum throughput. Moreover, for these two $\gamma$ values, it is beneficial in terms of throughput to allow FD operation (recall the discussion on Figs. \ref{fig:throughputs06} and \ref{fig:throughputs25} in the previous subsection).

Fig. \ref{fig:probRX_02} presents the values of $\Prx$ that maximize throughput versus the number of users when $\gamma=0.2$. We observe that in HD mode (see Fig. \ref{fig:probRX_02}(a)) and when the number of users exceeds a relatively high number (e.g., over $30$), the relay's receiver should be kept almost completely off to maximize throughput.
As discussed in the previous subsection, such a low $\gamma$ value yields a lot of successful user transmissions, and deactivating the relay ensures it's queue stability. We observe that the transmitter is not completely off at the same time (Fig. \ref{fig:probTX_02}(a)), which means that the relay still contributes to the network-wide throughput with rate being equal to the arrival rate, since the queue is stable. As a result, significant energy benefits are realized.

Similarly, when the number of users exceeds a relatively small number and the relay operates in FD mode, the receiver should be activated very infrequently (see Fig. \ref{fig:probRX_02}(b)). The transmitter should also be inactive under this users' load, but for a high number of users (see Fig. \ref{fig:probTX_02}(b)), its full cooperation ensures throughput maximization while the receiver's deactivation contributes considerably to energy savings.

\section{Summary and Conclusions}\label{sec:conclusion}

In this work, we have studied a full-duplex cooperative relay node, with capabilities to switch on or off the relay's receiver and transmitter independently, with the intention of assisting users in delivering their packets to one destination node. Multiple packet reception is assumed for both the relay and the destination node. The purpose is to give insights into how to set up a full-duplex relay node in order to maximize throughput and save network resources, and, thus, reducing energy consumption.   

We provide analytical expressions for the performance of the relay queue. More specifically, we derive equations for the per-user and network-wide throughput, the average arrival and service rate, conditions for the stability of the queue, and the average queue size. 
Additionally, we present the mathematical formulation for optimizing the values at which the receiver and the transmitter of the relay should be operating to maximize throughput, whilst queue stability is maintained at the same time. 

Our numerical results for the optimal values for which the relay's receiver and transmitter activation probabilities maximize throughput, as long as the relay's queue is stable, demonstrate the impact of the self-interference coefficient $g$ on the per-user and network-wide throughput. 
We showed that when the SINR threshold $\gamma$ is low and the relay operates in FD mode, it is beneficial - concerning energy efficiency - to switch off both the relay's receiver and transmitter when a relatively moderate amount of users is transmitting. Under the same $\gamma$, the relay's receiver should be deactivated when HD mode is used and a relatively high number of users is transmitting. 

\appendices
\section{Proof of Proposition \ref{thm:two-users}: Two asymmetric users performance analysis}\label{appxA} 

In this appendix, we prove Proposition \ref{thm:two-users}, which presents the relay queue performance for the two asymmetric users case.

We omit the calculations for the probability that the relay queue increases by $i$ packets when it is empty, $p^0_i$, or not empty, $p^1_i$, the probability that the queue decreases by one packet, $p^{1}_{-1}$, and the probability that the relay receives $i$ packets, $r^0_i$ when its queue is empty, or not empty, $r^1_i$. One can obtain these expressions following the methodology in Appendix A of \cite{Pappas_TWC_2015}. 

\subsection{Conditions for the Stability of the Queue}
The average service rate, $\mu$, is given by (\ref{eq:srvRate_N_users}).
The average arrival rate is given by (\ref{eq:arrRate_N_users}): 
\[ \lambda(\Prx, \Ptx) = P(Q=0)\lambda_0 + P(Q>0)\lambda_1,\]
where $\lambda_0 = \sum_{k=1}^{n} k r^0_k $ is the average arrival rate at the relay queue when the queue is empty, and $\lambda_1 = \sum_{k=1}^{n} k r^1_k  $ otherwise. 

According to Loyne's criterion \cite{Loynes-Criterion-1962}, a queue is stable if and only if the average arrival rate is strictly less than the average service rate. Hence:
 
\[ \lambda_1 < \mu \iff r^{1}_{1} + 2 r^{1}_{2} < \mu, \] \\where $r^{1}_{1} = (1-q_0\Ptx)A_1 + q_0\Ptx B_1,$ $r^{1}_{2} = (1-q_0\Ptx)A_2 + q_0\Ptx B_2, $ and $\mu = q_0 \Ptx A$.

The expressions for $A_1, B_1, A_2,B_2$ and $A$ can be calculated following the same methodology as in Appendix A of \cite{Pappas_TWC_2015}.

As a result, the values of $q_0$ for which the relay's queue is stable satisfies: $q_{0min} < q_0 < 1$, where:
\begin{equation}
	q_{0min} = \frac{A_1 + 2A_2}{ \Ptx A + A_1 + 2A_2 - B_1 - 2B_2}. 	
\end{equation}

\subsection{Average Queue Size}
The average queue size is known to be: $\bar{Q} = -S'(1)$, where $S'(1) = s_0 \frac{K''(1)}{L''(1)}$. 

The expressions for $K(z)$ and  $L(z)$ are given by:
\begin{equation} \begin{aligned}
K(z)= (-z^{-2}A(z) + z^{-1}A'(z) - B'(z))(z^{-1} -B(z))\\
-(z^{-1}A(z)-B(z))(-z^{-2}-B'(z)),
\end{aligned}
\end{equation}

\begin{equation}
L(z) = (z^{-1}-B(z))^2,
\end{equation}

where: 
$ A(z) = \sum_{i=0}^{n} a_i z^{-i},$ and $B(z) = \sum_{i=0}^{n+1} b_i z^{-i} $.

Then, $K''(1)$ and $L''(1)$ are given by:

\begin{multline}\label{K1}
K''(1)= (2A(1) - 2A'(1) + A''(1) - B''(1)) (-1-B'(1))\\
-(2 - B''(1))(-A(1) + A'(1) - B'(1)),
\end{multline}
\begin{multline}\label{L1}
L''(z) = [2(z^{-1}-B(z))(-z^{-2}-B'(z))]'\\
\Rightarrow L''(1)=2(-1-B'(1))^2.
\end{multline}

Therefore, the values of $A''(1)$ and $B''(1)$ are:

\begin{equation} A''(1) = 2p^0_1 + 6p^0_2 \end{equation} 
\begin{equation} B''(1) = 2-2p^1_{-1} + 4p^1_1+ 10p^1_2. \end{equation} 

Thus, the average queue size is given by (\ref{eq:avgQueueSz_2users}).

\section{Proof of Proposition \ref{thm:n-users}: Symmetric Users Performance Analysis}\label{appxB} 
In this appendix, we prove Proposition \ref{thm:n-users}, which presents
the relay queue performance for the symmetric users case.

We omit the calculations for the probability that the relay queue increases by $k$ packets when it is empty, $p^0_k$, or not empty, $p^1_k$, the probability that the queue decreases by one packet, $p^{1}_{-1}$, and the probability that the relay receives $k$ packets, $r^0_k$ when its queue is empty, or not empty, $r^1_k$. 
One can obtain these expressions following the methodology in Appendix B of \cite{Pappas_TWC_2015}. 

\subsection{Conditions for the Stability of the Queue}
The average service rate, $\mu$, is given by (\ref{eq:srvRate_N_users}).
The average arrival rate is given by (\ref{eq:arrRate_N_users}): 
\[ \lambda(\Prx, \Ptx) = P(Q=0)\lambda_0 + P(Q>0)\lambda_1,\]
where $\lambda_0 = \sum_{k=1}^{n} k r^0_k $ is the average arrival rate at the relay queue when the queue is empty, and $\lambda_1 = \sum_{k=1}^{n} k r^1_k  $ otherwise. 
The probability the relay receives $k$ packets, when its queue is empty, $r^{0}_{k}$, can be calculated using the methodology presented in Appendix B of \cite{Pappas_TWC_2015}. 

The probability the relay receives $k$ packets, when its queue is not empty,  is:
\begin{equation}
\begin{aligned}
r^{1}_{k} = (1-q_0 \Ptx) \sum\limits_{i=k}^{n} \binom {n}{i} \binom {i}{k}  q^i (1-q)^{n-i} P_{0,i,0}^{k} \times \\ 
\times (1-P_{d,i,0})^k [1-P_{0,i,0} (1-P_{d,i,0})]^{i-k} + \\
+ q_0 \Ptx \sum\limits_{i=k}^{n} \binom {n}{i} \binom {i}{k}  q^i (1-q)^{n-i} P_{0,i,1}^{k} (1-P_{d,i,1})^k \times \\
\times [1-P_{0,i,1} (1-P_{d,i,1})]^{i-k}, ~ 1 \leq k \leq  n.
\end{aligned} 
\end{equation}

The elements of the transition matrix are given by $a_k = p_k^0, b_0 = p_{-1}^{1}, b_1 = p_0^1$ and $b_{k+1}=p_k^{1}, \forall k>0$, where:
\begin{equation}
\begin{aligned}
p^0_k = \sum\limits_{i=k}^{n} \binom{n}{i} \binom{i}{k} q^i (1-q)^{n-i} P^k_{0,i,0} (1-P_{d,i,0})^k \times \\
\times [1-P_{0,i,0}(1-P_{d,i,0})]^{i-k}, ~ 1 \leq k \leq n,
\end{aligned} 
\end{equation}

\begin{equation}
\begin{aligned}
p^1_{-1} = q_0 \Ptx \sum\limits_{k=0}^{n} \binom{n}{k} q^k  (1-q)^{n-k}P_{0d,k} \times \\
\times [1-P_{0,k,1}(1-P_{d,k,i})]^{k},
\end{aligned} 
\end{equation}

\begin{equation}
\begin{aligned}
p^1_k = (1-q_0 \Ptx) \sum\limits_{i=k}^{n} \binom {n}{i} \binom {i}{k}  q^i (1-q)^{n-i} P^k_{0,i,0} \times \\
\times (1-P_{d,i,0})^k [1-P_{0,i,0}(1-P_{d,i,0})]^{i-k} + \\
+ q_0 \Ptx \sum\limits_{i=k}^{n} \binom {n}{i} \binom {i}{k}  q^i (1-q)^{n-i} (1-P_{0d,i}) P^k_{0,i,1} \times \\
\times (1-P_{d,i,1})^k [1-P_{0,i,1}(1-P_{d,i,1})]^{i-k} + \\
+ q_0 \Ptx \sum\limits_{i=k}^{n} \binom {n}{i} \binom {i}{k+1}  q^i (1-q)^{n-i} (1-P_{0d,i}) P^{k+1}_{0,i,1} \times \\
\times( 1-P_{d,i,1})^{k+1} [1-P_{0,i,1}(1-P_{d,i,1})]^{i-k-1}, ~ k \geq 1,
\end{aligned} 
\end{equation}

\begin{equation}
\begin{aligned}
p^1_0 = 1 - p^1_{-1} - \sum\limits_{i=1}^{n}p^1_i .
\end{aligned} 
\end{equation}

The probability that the queue in the relay is empty is given by \cite{Gebali:2015}:
\begin{equation} 
P(Q=0) = \frac{ 1 + B'(1) }{ 1 + B'(1) - A'(1) },
\end{equation}
where the expressions for $A'(1)$ and $B'(1)$ are

\begin{equation} A'(1) = -\lambda_0, \end{equation}
\begin{equation} 
B'(1) = -1 + p^1_{-1} - \sum\limits_{i=1}^{n} i p^1_i,
\end{equation}
since $ A(z) = \sum_{i=0}^{n} a_i z^{-i},$ and $B(z) = \sum_{i=0}^{n+1} b_i z^{-i} $. As a result, the \textit{probability that the queue in the relay is empty} is given by (\ref{eq:probRelayEmpty_Nusers}).

As in Appendix \ref{appxA}, the queue is stable of $\lambda_1 < \mu  \iff \sum_{k=1}^{n} k r^1_k < \mu$, where $r^1_k = (1-q_0\Ptx)A_k + q_0\Ptx B_k$ and $\mu = q_0 \Ptx A$. The expressions for $A, A_k, B_k$ are given in Appendix B of \cite{Pappas_TWC_2015}.

Thus, the values of $q_0$ for which the relay's queue is stable satisfies $q_{0min} < q_0 < 1$, where:
 
\begin{equation} 
q_{0min} = \frac{\sum\limits_{k=1}^{n} kA_k}{A \Ptx+\sum\limits_{k=1}^{n} k(A_k - B_k)}
\end{equation}

\subsection{Average Queue Size}
Similarly to the two asymmetric users case (see Appendix \ref{appxA}), the average queue size is given by $\bar{Q} = -S'(1)$, where $S'(1) = s_0 \frac{K''(1)}{L''(1)}$. 
The expressions for $K''(1)$ and $L''(1)$ are given by (\ref{K1}) and (\ref{L1}).
The expressions for $A''(1)$ and $B''(1)$ are:
\begin{equation} A''(1) = \sum\limits_{i=1}^{n} i(i+1)p^0_i,  \end{equation}
\begin{equation} B''(1) = 2-2p^1_{-1}+\sum\limits_{i=1}^{n} i(i+3)p^1_i. \end{equation}

Therefore, the \textit{average queue size} is given by (\ref{eq:avgQueueSz_Nusers}). 

\bibliographystyle{ieeetr}
\bibliography{paper}{}

\begin{thebibliography}{10}

\bibitem{Kramer-Cooperative-Communications-2006}
G.~Kramer, I.~Mari\'{c}, and R.~D. Yates, ``Cooperative communications,'' {\em
  {Foundations and Trends in Networking}}, vol.~1, pp.~271--425, Aug. 2006.

\bibitem{Rong-WiOpt2009}
B.~Rong and A.~Ephremides, ``Protocol-level cooperation in wireless networks:
  Stable throughput and delay analysis,'' in {\em 7th International Symposium
  on Modeling and Optimization in Mobile, Ad Hoc, and Wireless Networks
  (WiOPT)}, pp.~1--10, Jun. 2009.

\bibitem{Rong-ISIT-2009}
B.~Rong and A.~Ephremides, ``Cooperation above the physical layer: The case of
  a simple network,'' in {\em IEEE International Symposium on Information
  Theory (ISIT)}, pp.~1789--1793, Jun. 2009.

\bibitem{Pappas_TWC_2015}
N.~Pappas, M.~Kountouris, A.~Ephremides, and A.~Traganitis, ``Relay-assisted
  multiple access with full-duplex multi-packet reception,'' {\em IEEE
  Transactions on Wireless Communications}, vol.~14, pp.~3544--3558, Jul. 2015.

\bibitem{6231296}
K.~Zheng, F.~Hu, W.~Wang, W.~Xiang, and M.~Dohler, ``{Radio resource allocation
  in LTE-advanced cellular networks with M2M communications},'' {\em IEEE
  Communications Magazine}, vol.~50, pp.~184--192, Jul. 2012.

\bibitem{7248779}
G.~Rigazzi, N.~K. Pratas, P.~Popovski, and R.~Fantacci, ``{Aggregation and
  trunking of M2M traffic via D2D connections},'' in {\em IEEE International
  Conference on Communications (ICC)}, pp.~2973--2978, Jun. 2015.

\bibitem{IoT-Journal-2015-nrg-efficiency-D2D-res-alloc}
Z.~Zhou, M.~Dong, K.~Ota, G.~Wang, and L.~T. Yang, ``Energy-efficient resource
  allocation for {D2D} communications underlaying cloud-ran-based {LTE-A}
  networks,'' {\em IEEE Internet of Things Journal}, vol.~3, pp.~428--438, Jun.
  2016.

\bibitem{Systems-journal-2015-enabling-tech}
F.~K. Shaikh, S.~Zeadally, and E.~Exposito, ``Enabling technologies for green
  internet of things,'' {\em to appear in IEEE Systems Journal}.

\bibitem{IoT-Journal-2014-time-reversal-green}
Y.~Chen, F.~Han, Y.~H. Yang, H.~Ma, Y.~Han, C.~Jiang, H.~Q. Lai, D.~Claffey,
  Z.~Safar, and K.~J.~R. Liu, ``Time-reversal wireless paradigm for green
  internet of things: An overview,'' {\em IEEE Internet of Things Journal},
  vol.~1, pp.~81--98, Feb. 2014.

\bibitem{relay-benefits-2004}
J.~N. Laneman, D.~N.~C. Tse, and G.~W. Wornell, ``Cooperative diversity in
  wireless networks: Efficient protocols and outage behavior,'' {\em IEEE
  Transactions on Information Theory}, vol.~50, pp.~3062--3080, Dec. 2004.

\bibitem{FD-Relay-Survey-2015}
G.~Liu, F.~R. Yu, H.~Ji, V.~C.~M. Leung, and X.~Li, ``In-band full-duplex
  relaying: A survey, research issues and challenges,'' {\em IEEE
  Communications Surveys Tutorials}, vol.~17, pp.~500--524, Second quarter
  2015.

\bibitem{5G-Full-Duplex-Apps-2015}
H.~Alves, R.~D. Souza, and M.~E. Pellenz, ``{Brief survey on full-duplex
  relaying and its applications on 5G},'' in {\em IEEE 20th International
  Workshop on Computer Aided Modelling and Design of Communication Links and
  Networks (CAMAD)}, pp.~17--21, Sept. 2015.

\bibitem{5G-Survey-2015}
A.~Gupta and R.~K. Jha, ``A survey of {5G} network: Architecture and emerging
  technologies,'' {\em IEEE Access}, vol.~3, pp.~1206--1232, 2015.

\bibitem{5G-Self-Interference-Cancellation-2014}
S.~Hong, J.~Brand, J.~I. Choi, M.~Jain, J.~Mehlman, S.~Katti, and P.~Levis,
  ``{Applications of self-interference cancellation in 5G and beyond},'' {\em
  IEEE Communications Magazine}, vol.~52, pp.~114--121, Feb. 2014.

\bibitem{OnDynamicPolicies-Friderikos-ICC2012}
D.~Quintas and V.~Friderikos, ``On dynamic policies to switch off relay
  nodes,'' in {\em IEEE International Conference on Communications (ICC)},
  pp.~2945--2950, Jun. 2012.

\bibitem{Optimal-Relay-ICC-2011}
S.~Zhou, A.~J. Goldsmith, and Z.~Niu, ``On optimal relay placement and sleep
  control to improve energy efficiency in cellular networks,'' in {\em IEEE
  International Conference on Communications (ICC)}, pp.~1--6, Jun. 2011.

\bibitem{Sadek-2009}
A.~K. Sadek, K.~J.~R. Liu, and A.~Ephremides, ``Cognitive multiple access via
  cooperation: Protocol design and performance analysis,'' {\em IEEE
  Transactions on Information Theory}, vol.~53, pp.~3677--3696, Oct. 2007.

\bibitem{Simeone-TCOMM-2007}
O.~Simeone, Y.~Bar-Ness, and U.~Spagnolini, ``Stable throughput of cognitive
  radios with and without relaying capability,'' {\em IEEE Transactions on
  Communications}, vol.~55, pp.~2351--2360, Dec. 2007.

\bibitem{Pappas-Globecom-2010}
N.~Pappas, A.~Traganitis, and A.~Ephremides, ``Stability and performance issues
  of a relay assisted multiple access scheme,'' in {\em IEEE Global
  Telecommunications Conference (GLOBECOM 2010)}, pp.~1--5, Dec. 2010.

\bibitem{Pappas-ITW-2013}
N.~Pappas, M.~Kountouris, A.~Ephremides, and A.~Traganitis, ``On the stability
  region of a relay-assisted multiple access scheme,'' in {\em Information
  Theory Workshop (ITW), 2013 IEEE}, pp.~1--8, Sept. 2013.

\bibitem{Pappas-ISIT-2012}
N.~Pappas, J.~Jeon, A.~Ephremides, and A.~Traganitis, ``Wireless network-level
  partial relay cooperation,'' in {\em IEEE International Symposium on
  Information Theory Proceedings (ISIT)}, pp.~1122--1126, Jul. 2012.

\bibitem{WeeraddanaWiOpt2010}
C.~Weeraddana, M.~Codreanu, M.~Latva-aho, and A.~Ephremides, ``Resource
  allocation for cross-layer utility maximization in multi-hop wireless
  networks in the presence of self interference,'' in {\em Proceedings of the
  8th International Symposium on Modeling and Optimization in Mobile, Ad Hoc
  and Wireless Networks (WiOpt)}, pp.~70--75, May 2010.

\bibitem{WeeraddanaITW2010}
P.~C. Weeraddana, M.~Codreanu, M.~Latva-aho, and A.~Ephremides, ``The benefits
  from simultaneous transmission and reception in wireless networks,'' in {\em
  IEEE Information Theory Workshop (ITW)}, pp.~1--5, Aug. 2010.

\bibitem{Tse-book-2005}
D.~Tse and P.~Viswanath, {\em {Fundamentals of Wireless Communication}}.
\newblock New York, NY, USA: Cambridge University Press, 2005.

\bibitem{Loynes-Criterion-1962}
R.~M. Loynes, ``The stability of a queue with non-independent inter-arrival and
  service times,'' {\em Mathematical Proceedings of the Cambridge Philosophical
  Society}, vol.~58, pp.~497--520, Jul. 1962.

\bibitem{Gebali:2015}
F.~Gebali, {\em Analysis of Computer Networks}.
\newblock Springer Publishing Company, Incorporated, 2nd~ed., 2015.

\end{thebibliography}

\end{document}